\documentclass[pra,twocolumn,showpacs,preprintnumbers,amsmath,amssymb]{revtex4}
\usepackage{graphicx}
\usepackage{dcolumn}
\usepackage{bm}
\begin{document}

\title{Mean-field phase diagram of the 1-D Bose gas in a disorder potential}

\author{Luca Fontanesi}
\email{luca.fontanesi@epfl.ch}
\affiliation{Institute of Theoretical Physics, Ecole Polytechnique F\'ed\'erale de Lausanne EPFL, CH-1015 Lausanne, Switzerland}
\author{Michiel Wouters}
\affiliation{Institute of Theoretical Physics, Ecole Polytechnique F\'ed\'erale de Lausanne EPFL, CH-1015 Lausanne, Switzerland}
\author{Vincenzo Savona}
\affiliation{Institute of Theoretical Physics, Ecole Polytechnique F\'ed\'erale de Lausanne EPFL, CH-1015 Lausanne, Switzerland}


\begin{abstract}
We study the quantum phase transition of the 1D weakly interacting Bose gas in the presence of disorder. We characterize the phase transition as a function of disorder and interaction strengths, by inspecting the long-range behavior of the one-body density matrix as well as the drop in the superfluid fraction. We focus on the properties of the low-energy Bogoliubov excitations that drive the phase transition, and find that the transition to the insulator state is marked by a diverging density of states and a localization length that diverges as a power-law with power $1$. We draw the phase diagram and we observe that the boundary between the superfluid and the Bose glass phase is characterized by two different algebraic relations. These can be explained analytically by considering the limiting cases of zero and infinite disorder correlation length.
\end{abstract}

\maketitle

\section{Introduction}
The effect of disorder on quantum systems is a subject of both fundamental and practical interest. Since the seminal work of Anderson \cite{anderson_localization}, it has become clear that what seems at first sight a nuisance is in fact a source of very rich physical behavior: the disorder is not just a perturbation of the wave functions, but localizes the low energy states in three dimensions; in lower dimensions even all single particle states are localized.

A single particle picture can be a good approximation of a dilute gas, but in practice interactions between particles play often a crucial role. Understanding of the interplay of disorder and interactions is therefore of fundamental importance. In the context of solid state physics, the fermionic problem is the most relevant one, but thanks to the enormous progress in experimental control over ultracold atomic gases, the problem of the disordered Bose gas has become a subject of a vigorous research activity as well. Anderson localization has been experimentally observed recently in one dimensional bosonic systems with vanishing interaction~\cite{aspect_anderson,inguscio_anderson}. More recent experiments have moved away from the limit of vanishing interactions, so to study the localized (Bose glass) to superfluid phase transition~\cite{inguscio_delocalization,hulet}.

The theoretical interest in this phase transition dates back much longer and a variety of theoretical techniques have been used to tackle the problem. There are two main regimes that have been considered, one marked by weak disorder and arbitrary interactions, the other characterized by weak interactions and arbitrary disorder. The former has been the object of the first investigations in the eighties. Using a renormalization group analysis, Giamarchi and Schultz~\cite{giamarchi} were able to study the quantum phase transition in the limit of weak disorder in one dimension. The picture emerging from this analysis was that for a finite amount of disorder, a minimal strength of interactions is required to break the Anderson localization, but that for too strong interactions, the system is driven into a strongly correlated localized Bose gas phase. Their renormalization group approach was able to study the latter phase transition quantitatively, but it could not be clarified whether the transition at the weak interaction side is of the same nature. The interplay between periodic and disordered potentials was first addressed in the seminal work by Fisher {\em et al.}~\cite{fisher_superfluidinsulator}, where the insulating disordered phase, named `Bose glass', was contrasted to the Mott insulator phase by its compressible nature and to the superfluid phase by its vanishing superfluid stiffness. Quantum Monte Carlo~\cite{prokofev_comment} and density matrix renormalization group~\cite{rapsch} studies have investigated in detail the disorder phase diagram in the limit of strong interactions close to the Mott insulator phase.

Early experimental efforts to reach the Bose glass phase coming from the Mott insulating phase were presented in Ref.~\cite{inguscio_boseglass}. More recently several experimental works have addressed the quite different regime where the gas is in the weakly interacting limit and with many particles in each potential minimum. Above a critical strength of the disorder, fragmentation of the condensate density and loss of spatial coherence were identified~\cite{inguscio_delocalization,hulet}. All these experiments have considered the one dimensional geometry. We will restrict our analysis to this case as well. Thanks to the weak interaction limit, the theoretical description of these experiments can be performed in a first approximation with the Gross-Pitaevskii equation. The Gross-Pitaevskii equation in the presence of a disorder potential was recently studied by several groups~\cite{falco,lugan_phase,gurarie,pavloff}. For the condensate wave function, two different cases were identified: a connected density profile at weak and a fragmented one at strong disorder.
The properties of the elementary Bogoliubov excitations stemming from the Gross-Pitaevskii state were studied analytically in the superfluid regime. Their localization length was shown to exhibit a power law behavior as a function of energy: $E^\alpha$, with $\alpha=2$ deep in the superfluid phase~\cite{gurarie} and $\alpha=1$ at the phase transition~\cite{pavloff}. Numerically we found that $\alpha<1$ in the Bose glass phase~\cite{fontanesi}. The density of states (DOS) was shown to be constant in the superfluid phase as for phonons in random chains~\cite{ziman}; in the Bose glass phase, a divergence of the low energy DOS was numerically identified in our previous work. This result is in accordance with the real space renormalization group analysis by Altman {\em et al.}~\cite{altman_09} with particle hole symmetry. Fisher {\em et al.}~\cite{fisher_superfluidinsulator} on the other hand argued that the low energy DOS should be constant in the Bose glass phase. This difference in behavior could be due to a different nature of the phase transition between the superfluid and glassy phase for weak and strong interaction as suggested by Giamarchi and Schultz~\cite{giamarchi}.

In the present work, we make a deeper analysis of the condensate wave function and Bogoliubov excitations in the different phases. We carry out a new analysis of the superfluid fraction across the phase transition. In Ref.~\cite{fontanesi}, the phase diagram was restricted to the limit of weak interaction energy where the healing length $\xi$ was large as compared to the correlation length of the disorder $\eta$. In that case, the disorder is effectively an uncorrelated white noise (WN) potential. In this work, we extend the phase diagram to the regime where the $\eta \gg \xi$. Kinetic energy is then not important to determine the density profile of the condensate and the Thomas-Fermi (TF) approximation is accurate. The WN and TF regimes are marked by two different power-law relations between interaction and disorder at the phase boundary. For these we provide a rigorous analytical arguments.

The paper is organized as follows. In Sec. \ref{sec_theory} the theoretical model is presented in detail. Sec. \ref{sec_phase} is devoted to the predictions of the extended mean-field model. In Sec. \ref{sec_coherence} the study of the correlation length is presented. An analysis of the density of states and localization properties of the Bogoliubov excitations is carried out in Sec. \ref{sec_DOS}. In Sec. \ref{sec_superfluid} a study of the superfluid fraction of the gas is performed. The phase diagram is discussed in Sec. \ref{sec_pd}. Our conclusions are presented in Sec. \ref{section_conclusions}.

\section{Theory}\label{sec_theory}

We want to study the properties of the 1D Bose gas at zero temperature. These gases are well described by the mean field theory in 3D, but in lower dimensions no real condensate is present and the standard mean field theory is no longer valid. Nevetheless, the Bogoliubov approach can be extended to weakly interacting low-dimensional bosonic systems in a density-phase version defined on a lattice~\cite{mora}. A homogeneous 1D Bose gas is in the weakly interacting regime when $\rho\xi \gg 1$, where $\xi=\hbar/\sqrt{m\rho g}$ is the healing length, $\rho$ the total density, $g$ the coupling constant and $m$ the mass.

The Hamiltonian describing the bosonic system is
\begin{eqnarray}\label{hamiltonian}
\hat H  =  \int \mathrm{d}{r} \; \left[\hat{\Psi}^\dagger({r}) \hat H_0  \hat{\Psi}({r}) + \frac{g}{2}\hat{\Psi}^\dagger({r})\hat{\Psi}^\dagger({r})\hat{\Psi}({r})\hat{\Psi}({r})\right],
\end{eqnarray}
where $\hat{H}_0=-\hbar^2\partial_r^2/(2m)+V({r})$ is the single-particle Hamiltonian, $\hat\Psi$ is the field operator and $V(r)$ the external potential. Here we study the case where $V(r)$ is a Gauss-distributed and Gauss-correlated disorder:
\begin{equation}
\langle V({r})V({r'})\rangle=\Delta^2 e^{-\frac{(r-r')^2}{2\eta^2}},
\end{equation}
where $\Delta$ is the disorder amplitude and $\eta$ is the spatial correlation length. We make the general choice of Gauss-distributed disorder.
We are aware that many experiments aimed at the characterization of the phase transition deal with speckle potentials~\cite{hulet,aspect_anderson}. These potentials have a lower bound and their statistical distribution does not show a lower Gaussian tail.
Our analysis could be easily extended to this case, as $V(r)$ is only treated numerically. On the other hand, the assumption of spatial Gauss-correlation is consistent with the experimental realizations~\cite{hulet}.

In low dimensionality, the Bogoliubov approach requires the definition of the field in terms of density $\hat\rho$ and phase $\hat\theta$ operators as $\hat\Psi=e^{i\hat\theta}\sqrt{\hat\rho}$. In the mean field approach one splits the density operator in a c-field part, $\rho_0$, and a fluctuation term $\delta \hat\rho$. A correct definition of the phase operator is only possible~\cite{mora} by introducing a spatial discretization of step size $\ell$. Provided that the system is in the high density regime, $\rho \ell = (\rho_0+\langle\delta\hat{\rho}\rangle) \ell> 1$, and that every cell is largely populated, the phase operator can be defined precisely. The density fluctuation can then be treated perturbatively: this perturbative approach is valid under the assumption of small density fluctuations, $\delta \hat \rho/\rho\ll 1$, and spatially slowly varying phase fluctuations, $\hat \theta_{i+1}-\hat \theta_{i} \ll 1$.

The ground state density profile $\rho_0$ obeys the Gross-Pitaevskii equation
\begin{equation}
\left[\hat{H}_0+g\rho_0({r})\right]\sqrt{\rho_0({r})}=\mu\sqrt{\rho_0({r})},
\label{GPE}
\end{equation}
where $\mu$ is the chemical potential. The excitations of the system can be computed via the Bogoliubov-de Gennes equations, obtained by linear expansion around the Gross-Pitaevskii solution
\begin{eqnarray}\label{BdG}
\left(\hat{H}_0+2g\rho_0({r})-\mu\right)u_j({r}) + g\rho_0({r}) v_j({r}) & = & E_j u_j({r}),\nonumber \\
- g\rho_0({r}) u_j({r}) - \left(\hat{H}_0+2g\rho_0({r})-\mu\right)v_j({r}) & = & E_j v_j({r}),\nonumber \\
\end{eqnarray}
that define the Bogoliubov $u_j({r})$ and $v_j({r})$ modes, normalized as
\begin{equation}
\int \mathrm{d}{r} \; |u_j(r)|^2-|v_j(r)|^2=1.
\end{equation}
Phase and density operators can be expressed in terms of these wave functions as~\cite{shevchenko}
\begin{eqnarray}
\hat\theta(r) & = & \frac{1}{2i\sqrt{\rho(r)}}\sum_j [\theta_j(r) \hat b_j - \theta_j^*(r)\hat b^\dagger],\nonumber \\
\delta\hat\rho(r) & = & \sqrt{\rho(r)}\sum_j [\rho_j(r) \hat b_j+ \rho_j^*(r)\hat b^\dagger],
\end{eqnarray}
where $b_j$ and $b_j^\dagger$ are bosonic annihilation and creator operators of an excitation with energy $E_j$ and with
\begin{equation}
u_j(r)=\frac{\rho_j(r)+\theta_j(r)}{2},\quad v_j(r)=\frac{\rho_j(r)-\theta_j(r)}{2}.
\end{equation}
The extended Bogoliubov approach requires an orthogonalization of these modes with respect to the ground state \cite{mora}, which finally brings to the new modes $u_{\perp j}({r})$ and $v_{\perp j}({r})$.

Taking advantage of Wick's theorem, the one-body density matrix at $T=0$, computed with the extended Bogoliubov method, takes the form
\begin{equation}\label{g1}
G({r},{r'}) = \sqrt{\rho({r})\rho({r'})}e^{-\frac{1}{2}\displaystyle\sum_j \left|\frac{v_{\perp j}({r})}{\sqrt{\rho_0({r})}} -\frac{v_{\perp j}({r'})}{\sqrt{\rho_0({r'})}}\right|^2},
\end{equation}
where the only contribution comes from the quantum fluctuations. This two-point function is not self-averaging. We will preferably work with the spatially averaged \emph{degree of coherence}
\begin{equation}\label{coherence}
g_{1}({r})=\frac{1}{L}\int \mathrm{d}{r'} \frac{G({r},{r'})}
{\sqrt{\rho({r})\rho({r'})}}.
\end{equation}
that directly gives information about the decay of the one-body density matrix. This quantity, in the quasi-condensed phase, is expected to show an algebraic decay~\cite{fisher_superfluidinsulator}, as in the homogeneous case~\cite{popov_g1}, together with a finite superfluid fraction. In presence of disorder the system can undergo a quantum phase transition to the Bose glass phase characterized by an exponential decay of the one-body density matrix~\cite{fisher_superfluidinsulator}.

\section{Ground State and phase fluctuations}\label{sec_phase}

Three energy scales enter our problem: the amplitude of the disorder $\Delta$, the interaction energy, $U=gN_0/L$, where $N_0$ is the number of particle in the ground state, and the energy associated to the correlation length of the potential $E_c=\frac{\hbar^2}{2m\eta^2}$. In the analysis that follows the problem is rescaled with respect to this latter.
We have performed numerical calculations on systems of finite size (up to $4096\,\eta$), adopting periodic boundary conditions for the equations and for the disordered potential $V(r)$. Every quantity has been averaged on several disorder configurations. By varying the system size, we could extract the limiting behavior of each quantity in the thermodynamic limit. In order to describe the system in the continuous limit we fulfill the condition that the kinetic hopping energy $t=\frac{\hbar^2}{2ml^2}$ be much larger than any other characteristic energy of the system. This must hold in particular for the energies $U$, $\Delta$, $\mu$ and $E_c$. In the numerical simulations that follow we took an average quasicondensate density of $N_0\eta/L=8$. It is important to remark that in the mean field limit the phase boundary, determined from the Bogoliubov-de Gennes equations and the functional shape of the degree of coherence, does not depend on the density and interaction strength separately, but only on the product of the two quantities that appears in Eqns. (\ref{GPE}) and (\ref{BdG}). As outlined in Sec. \ref{sec_theory}, this model is valid in the limit of large density and its prediction are increasingly accurate in the limit $\rho\to \infty$, $g\to 0$ at constant $g\rho$.

\begin{figure}
\includegraphics[width=0.5\textwidth]{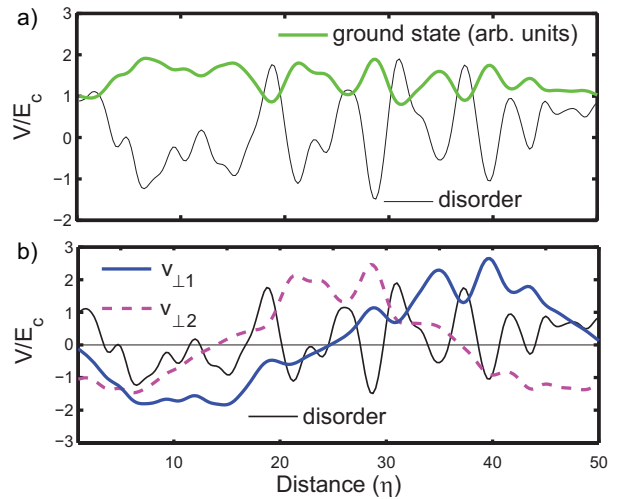}
\caption{(color online)  Quasicondensed phase: {\bf a.} ground state wave function and {\bf b.} first two excitations $v_\perp j$ for $U=1.44\,E_c$ and $\Delta=0.8\,E_c$.}
\label{fig_SF}
\end{figure}

\begin{figure}
\includegraphics[width=0.5\textwidth]{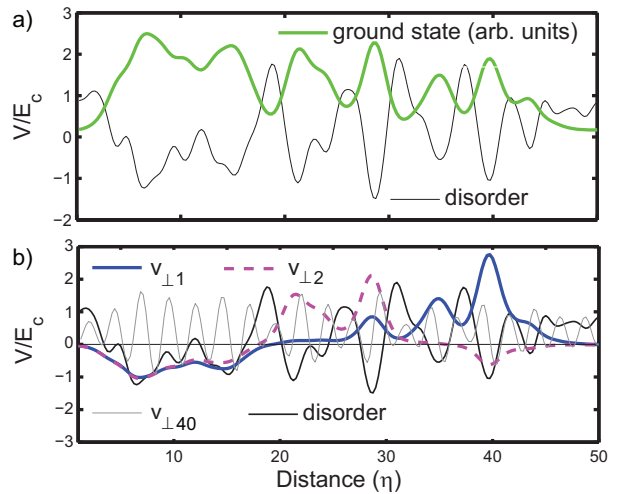}
\caption{(color online) Bose glass phase: {\bf a.} ground state wave function and {\bf b.} first two excitations $v_\perp j$ for $U=0.48\,E_c$ and $\Delta=0.8\,E_c$. The thin grey line represents an excitation at higher energy, namely $v_{\perp 40}({r})$.}
\label{fig_BG}
\end{figure}

In Fig. \ref{fig_SF}.a the ground state is shown for $\Delta=0.8\,E_c$ and $U=1.44\,E_c$, in the quasicondensed phase, computed via Eq. (\ref{GPE}). In Fig. \ref{fig_SF}.b the first two excitations $v_{\perp j}({r})$ appear ($j=1,2$). The same quantities computed in the Bose glass phase ($\Delta=0.8\,E_c$ and $U=0.48\,E_c$) are shown in Fig. \ref{fig_BG}.

It is evident, comparing Figs. \ref{fig_SF}.a and \ref{fig_BG}.a, that in the case of stronger interaction the ground state wave function is smoother. In fact the interaction term involves a more homogeneous effective potential in Eq. (\ref{GPE}) where the bare disorder potential is screened. In a case dominated by the disorder, the ground state separates into fragments linked by regions with an exponentially vanishing wave function, that can be seen as weak links.

Low energy excitations have a delocalized shape and follow the modulation of the ground state wave function, but, differently from the ground state, they show a phase character with a number of nodes that increases for increasing energy. In the homogeneous case the lowest energy excitations are found to be plane waves. The phase fluctuations preserve a plane wave profile in the quasicondensed phase, as can be seen in Fig. \ref{fig_SF}.b, only slightly modulated by the underlying disordered potential. On the other hand, at smaller values of $U$, in the Bose glass case, the disorder starts to compete with the interaction and the $v_{\perp j}({r})$ modes start losing their regular shape, developing nodes in correspondence with low density zones, i.e. with high barriers of the potential. This is shown in Fig. \ref{fig_BG}.b, together with an example of an excitation at higher energy that displays a fastly oscillating behavior and does not contribute in determining long-range properties.

\begin{figure}
\includegraphics[width=0.5\textwidth]{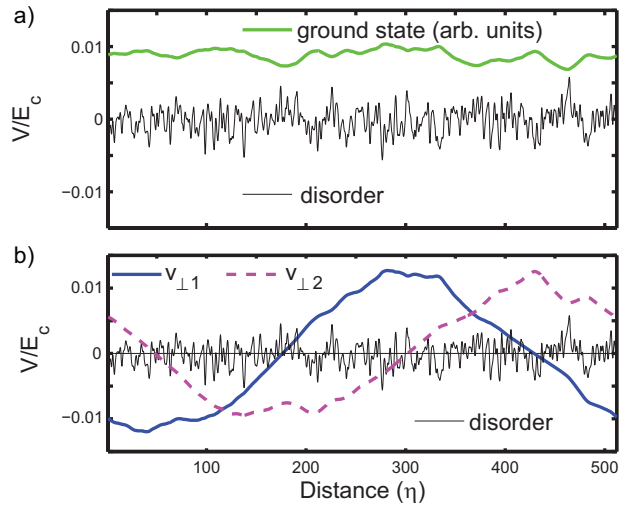}
\caption{(color online) {\bf a.} Ground state and {\bf b.} first two excitations $v_\perp j$ for $U=\Delta=1.6\times10^{-3}\,E_c$, where the system is in the superfluid phase and in the WN regime.}
\label{fig_WN}
\end{figure}

\begin{figure}
\includegraphics[width=0.5\textwidth]{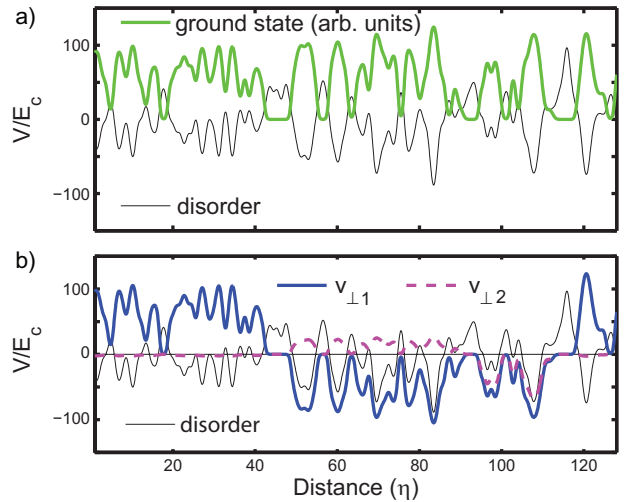}
\caption{(color online) {\bf a.} Ground state and {\bf b.} first two excitations $v_\perp j$ for $U=\Delta=25.6\,E_c$, where the system is in the Bose glass phase and in the TF regime.}
\label{fig_TF}
\end{figure}

It is instructive to analyze the role of the correlation length $\eta$ on the ground state and on the excitations. The quantity $E_c$ and its relation with $\mu$ turn out to be essential in determining the regime of the Bose gas. In fact small values of $U/E_c$ and $\Delta/E_c$ imply that $\eta \ll \xi$ and the ground state wavefunction is spread over many correlation lengths of the potential. In this limit, the disorder is equivalent to a WN potential. On the contrary, for large values of $U/E_c$ and $\Delta/E_c$, the system enters the TF regime, where the kinetic term is negligible and the ground state follows the spatial variations of the potential. In Fig. \ref{fig_WN}a the ground state is shown for a case close to the WN limit ($U=\Delta=1.6\times10^{-3}\,E_c$). The same quantity is shown in Fig. \ref{fig_TF}a for $U=\Delta=25.6\,E_c$, close to the TF regime. The differences explained above are evident in the distributions of the ground state that is spread over many correlation lengths in the WN case, whereas it fills the potential minima in the TF regime and it approaches the form
\begin{eqnarray}
|\phi_0(r)|^2 & = & \frac{\mu-V(r)}{UL}, \qquad  \textrm{for} \qquad \mu>V(r) \nonumber \\
|\phi_0(r)|^2 & = & 0, \qquad  \textrm{otherwise}.
\end{eqnarray}

The WN case presented in Fig. \ref{fig_WN} is in the superfluid phase. The low lying excitations are very close to plane waves with small modulations and a regular spacing between the nodes. The TF example (see Fig. \ref{fig_TF}) is in the Bose glass phase and the nodes of the excitations are pinned to the low condensate density regions. Their profile follows closely the condensate amplitude.

Note that the ratio between interaction and disorder amplitudes is in both cases $\Delta/U=1$. The transition between superfluid and Bose glass phases can apparently be tuned by only varying the disorder correlation length $\eta$. We will come back to this point in Sec. \ref{sec_pd} where we present the phase diagram.

\begin{figure}
\includegraphics[width=0.5\textwidth]{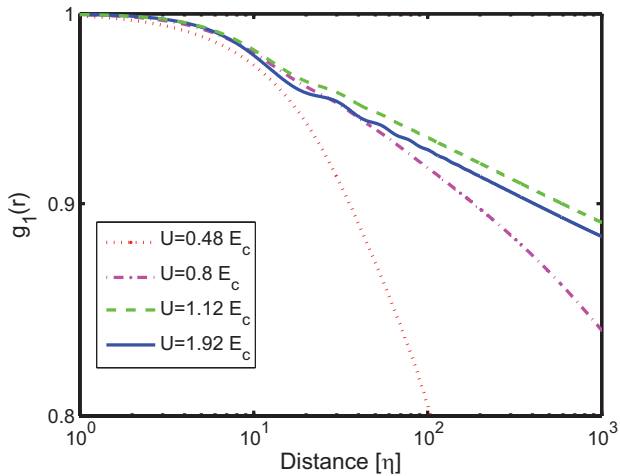}
\caption{(color online) Degree of coherence $g_1(r)$ for fixed value of disorder $\Delta=0.8\,E_c$ for different values of the interaction energy. The cases with $U=0.48-0.8\,E_c$ are in the Bose glass phase, whereas for $U=1.12-1.92\,E_c$ are superfluid.}
\label{fig_correlation}
\end{figure}

\section{Correlation length}\label{sec_coherence}

As outlined above, the phase transition is characterized by inspecting the long range behavior of the degree of coherence expressed by Eq. (\ref{coherence}). Fig. \ref{fig_correlation} shows the degree of coherence for the same value of disorder ($\Delta=0.8\,E_c$) varying the interaction energy. We show only the spatial interval $[0,L/4]$ that reflects the $L\to \infty$ behavior: at longer distance, deviations due to the periodic boundary conditions affect this quantity, as verified by studying the scaling with $L$. For the lowest value of the interaction energy, $g_1(r)$ shows an exponential decay that marks the Bose glass phase in a situation where the disorder breaks the long-range coherence. By increasing the interaction energy the decay becomes slower up to a certain value of interaction that drives the system in the quasicondensed phase characterized by an algebraic decay of $g_1(r)$ (linear in the double-logarithmic scale of Fig. \ref{fig_correlation}). By further increasing $U$, $g_1(r)$ still displays a power-law decay but falls off more rapidly as a function of $r$. This is in analogy with the homogeneous case~\cite{pitaevskii} where interactions cause quantum fluctuations and loss of coherence. At much larger values of the interaction energy this trend drives the system in the Tonk-Girardeau regime that cannot be described by the mean field theory. A similar analysis performed at fixed $U$ shows a monotonic reduction of the coherence when increasing the disorder strength.

\begin{figure}
\includegraphics[width=0.5\textwidth]{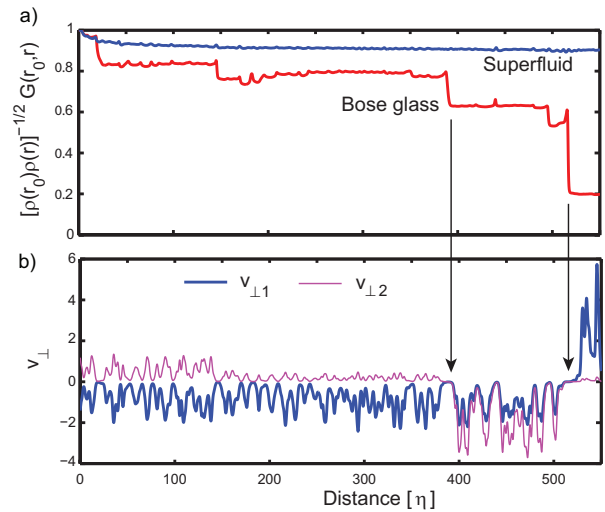}
\caption{(color online) {\bf a.} $G(r_0,r)$ in the superfluid and in the Bose glass phases. {\bf b.} corresponding low-energy $v_\perp$ excitations in the Bose glass phase. The arrows point out the link between the jumps in $G(r_0,r)$ and the nodes of the low-energy excitations.}
\label{fig_2points}
\end{figure}

To better understand the physics behind the functional dependence of $g_1(r)$, it is useful to focus on the 2-point correlation function expressed in Eq. (\ref{g1}). Fig. \ref{fig_2points} shows $G({r_0},{r})$ for two cases in the Bose glass and quasicondensate phases respectively. It is evident that the correlation in the superfluid phase, although slightly modulated by the underlying disorder profile, displays a smooth decay. On the other hand, the Bose glass case is characterized by abrupt jumps in the 2-point coherence that separate relatively coherent zones. It is immediate to notice that these jumps coincide with the nodes of the lowest energy excitations as can be checked comparing with the corresponding $v_{\perp j}(r)$-modes in Fig. \ref{fig_2points}b. The size of the jumps is related to the amplitude of the excitations, that increases for decreasing energy.

\section{Density of States and Localization}\label{sec_DOS}

Given that the loss of coherence in the Bose glass phase is due to low-lying Bogoliubov excitations, we study their properties in detail. In particular we are interested in the DOS and the localization properties of the $v_{\perp j}$-modes, that are the only ones playing a role at zero temperature.

\begin{figure}
\includegraphics[width=0.5\textwidth]{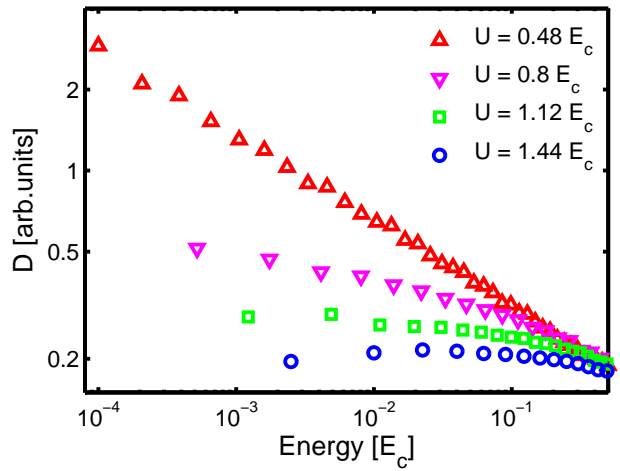}
\caption{(color online) Averaged $D(E)$ at fixed $\Delta=0.8\,E_c$ for various interaction energies. The triangles denote two Bose glass cases ($U=0.48-0.8\,E_c$), whereas squares and circles mark two superfluid phases ($U=1.12-1.44\,E_c$).}
\label{DOS}
\end{figure}
The DOS is defined as
\begin{equation}
D(E)=\sum_j \delta(E-E_j),
\end{equation}
where the $E_j$ are the positive Bogoliubov energies. This quantity is expected to approach a constant value for $E\to 0$ in the superfluid phase, as for phonons in random elastic chains~\cite{ziman}. Moreover, some theoretical studies~\cite{fisher_superfluidinsulator,zhang} have argued that the low-energy DOS should remain constant also in the Bose glass phase. The results of our mean field calculation disagree with this latter prediction in the mean field limit. As shown in Fig. \ref{DOS}, in the case dominated by interaction, $D(E)$ approaches a constant value at low energy, while it develops a power-law divergence in the Bose glass phase. In fact, by decreasing the interaction strength at fixed disorder, the DOS for $E\to0$ increases and it starts diverging following a power-law beyond the phase boundary. In addition, the slope of the power-law increases monotonically going deeper in the insulator phase.

We can connect this divergence to the behavior of $g_1$. Inspection of Eq. (\ref{g1}) shows that at long distances, the main contribution comes from the term $\sum_j|v_{\perp j}(r)|^2$. Thus, we can rewrite $G(r,r_0)\sim\exp[-\int |v_{E\perp}(r)|^2 D(E) \mathrm{d}E]$ (see also Ref.~\cite{zhang}). Here $|v_{E\perp}(r)|^2$ is defined as the local density of Bogoliubov excitations per unit energy, i.e.
\begin{equation}
 |v_{E\perp}(r)|^2=\frac{\sum_j|v_{\perp j}(r)|^2 \delta(E-E_j)}{\sum_j \delta(E-E_j)}.
\end{equation}
This quantity is expected to diverge as $1/E$, as it is the case both in the homogeneous system and in the case of weak links between junctions~\cite{leggett_josephson} (as also checked numerically for a disordered potential). Hence, the change in $g_1(r)$ can be linked to a change in $D(E)$, that changes from constant to a power-law divergence for $E\to 0$. These low energy excitations change the phase between weakly coupled neighbouring islands. The DOS is therefore directly related to the statistics of the strength of the weak links investigated in Ref.~\cite{altman_09}. Our diverging DOS is in agreement with their analysis.

The discrepancy between our results and former predictions~\cite{fisher_superfluidinsulator,zhang} could also be due to a different nature of the phase transition in the weakly and strongly interacting regimes, as also suggested by a recent renormalization group analysis~\cite{altman_09}. These results might imply two glassy phases characterized by different properties. In particular the analysis by Altman {\em et al.} has found a so-called \emph{random-singlet phase} characterized by a divergent DOS and claims that the phase transition at strong disorder belongs to a different universality class with respect to the weak disorder transition~\cite{giamarchi}. The random-singlet phase is specific of systems with particle-hole symmetry, as it is the case for the Bogoliubov model studied here.

\begin{figure}
\includegraphics[width=0.5\textwidth]{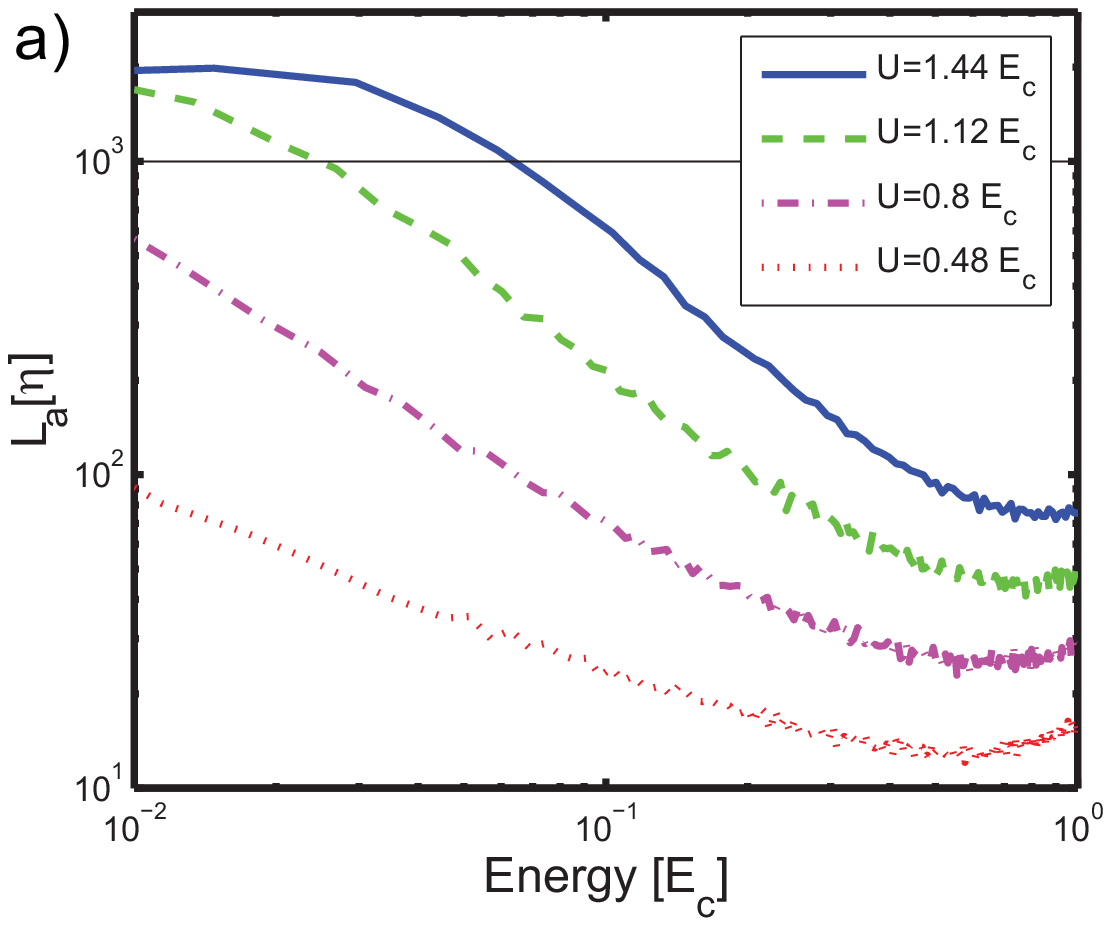}
\includegraphics[width=0.5\textwidth]{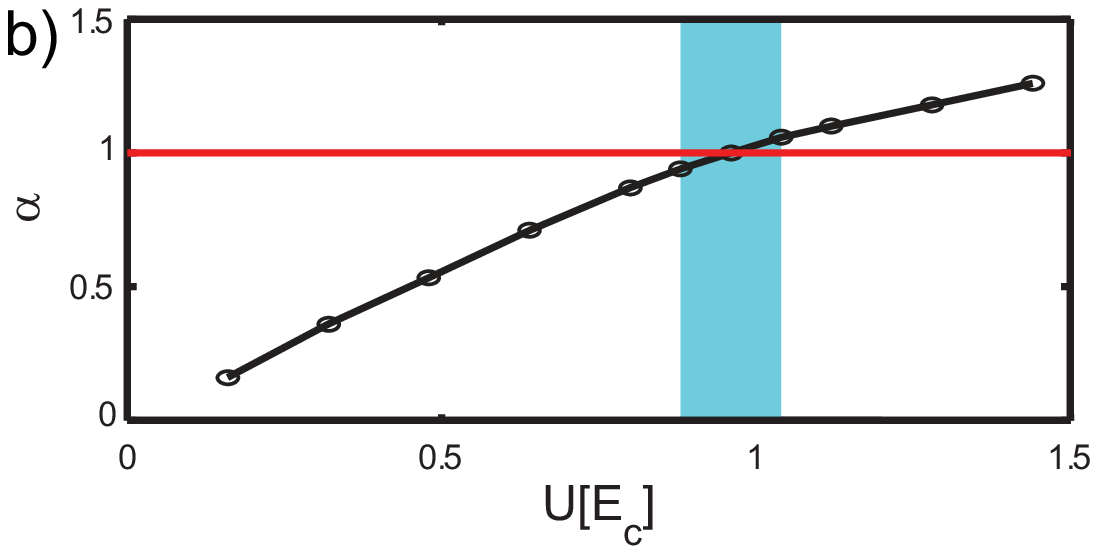}
\caption{(color online) {\bf a.} Averaged IPN for fixed $\Delta=0.8\,E_c$ varying $U$. The cases with $U=0.48-0.8\,E_c$ are in the Bose glass phase, whereas for $U=1.12-1.44\,E_c$ are superfluid. A saturation of the IPN occurs at large lengths because of the finte size of the simulated system, the horizontal black line represents a confidence limit. {\bf b.} Computed exponents of the power-law divergence of $L_a(E)$ for different values of $U$.}
\label{fig_ll}
\end{figure}

We now turn to the localization properties of the Bogoliubov modes. As pointed out in the previous sections, in presence of interaction, delocalized low-energy phase fluctuations are the main mechanism of reduction of coherence. The localization properties of these excitations is still under debate~\cite{gurarie,pavloff}. As a measure of localization we choose the inverse participation number (IPN), that directly gives an estimate of the spatial extent of the wave function. We notice however that IPN can be very different from other quantities characterizing the localization as, for instance, the exponential decay length of the wavefunction tails~\cite{kramer}. The IPN is nevertheless the most relevant characterization for our purposes. Indeed, a phase change over a long distance can only be produced by an excitation whose wave function is significantly non-zero at points very far apart in space (even if it has rapidly decaying exponential tails outside these regions~\cite{kramer}). The IPN is defined as
\begin{equation}
\frac{1}{I_j} = \frac{\int \mathrm{d}{r} |v_{\perp j}({r})|^4}{\left(\int \mathrm{d}{r}|v_{\perp j}({r})|^2\right)^2},
\end{equation}
and the corresponding realization-averaged quantity is $L_a(E)=\sum_j I_j \delta(E-E_j)/D(E)$. In Fig. \ref{fig_ll} the results are shown for fixed disorder and varying interaction strength. As it can be noticed, the IPN always shows a power-law divergence $E^{-\alpha}$ for $E\to0$, with $\alpha$ increasing when going deeper in the superfluid phase. The finite size of the simulation limits our analysis for large $U$, where the IPN saturates, as can be seen in Fig. \ref{fig_ll}.a. We draw a horizontal line to separate the region of the plot that is not affected by the finite size of the numerical sample.

As it is shown in Fig. \ref{fig_ll}.b, we find an exponent varying continuously around $1$ and crossing that value in correspondence with the boundary computed by the correlation length. In the quasicondensed phase $\alpha>1$, whereas we find $\alpha<1$ in the Bose glass phase. It decreases by lowering the interaction strength, apparently linearly vanishing for $U\to 0$ as predicted for the non interacting case, where the localization goes to a constant value at low energy. We can compare our results with a recent theoretical work~\cite{gurarie} that predicts an exponent $\alpha=1$ at the phase transition and $1<\alpha<2$ in the superfluid phase. We find a full agreement with this prediction, although it is not possible to reach the case with exponent $\alpha=2$ because of the finite size of our simulations.

A precise characterization of a phase boundary via the DOS turns out to be difficult, whereas the IPN analysis gives the boundary with high accuracy as evident in Fig. \ref{fig_ll}.b.

\section{Superfluid fraction}\label{sec_superfluid}

\begin{figure*}
\includegraphics[width=16cm]{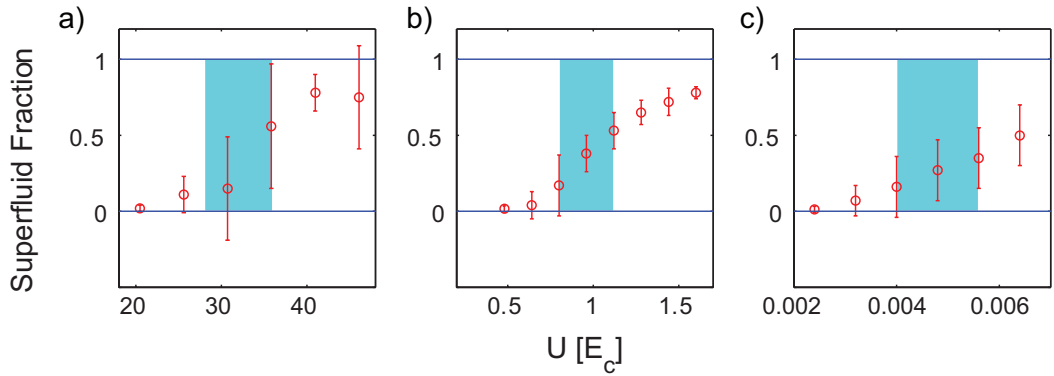}
\caption{(color online) Superfluid fraction for 3 values of disorder: {\bf a.} $\Delta=12.8\,E_c$, {\bf b.} $\Delta= 0.8\,E_c$, {\bf c.} $\Delta= 0.016\,E_c$. The average superfluid fractions and their error bars are shown as a function of the interaction energy. The shaded zones mark the phase transition computed through the degree of coherence.\label{fig_superfluid}}
\end{figure*}

An alternative way to characterize the thermodynamic phase of the gas is the computation of the superfluid fraction. The usual approach to superfluidity is the two-fluid picture and the distinction between normal fluid and superfluid resides in the different response to a small velocity field. This turns out to be equivalent to imposing a phase twist in the boundary conditions~\cite{lieb_2002}. The superfluid fraction can be computed from the energy difference between the system at rest and the moving one, namely
\begin{equation}
f_S=\frac{2mL^2}{\hbar^2N}\lim_{\Theta\to 0}\frac{E_\Theta-E_0}{\Theta^2},
\end{equation}
where $\Theta$ is the total phase twist. $E_\Theta$ is the energy of a condensate with twisted boundary conditions ($\Psi(L)=\Psi(0)e^{i\Theta}$) and $E_0$ is the ground state energy of the system at rest. We consider a small phase twist, ($\Theta=\pi/32\ll \pi$), to avoid excitations and level crossing. With a gauge transformation
$\Psi(x)\to\tilde\Psi e^{i\Theta x /L}$, the twisted boundary problem is mapped on a problem with periodic boundary conditions, with shifted momentum $p\to p+\hbar\Theta/L$, so that $\nabla \rightarrow \nabla+i\Theta/L$. This substitution enters both the Gross-Pitaevskii equation (\ref{GPE}) and the Bogoliubov-de Gennes equations (\ref{BdG}). In the homogeneous case, Galilean invariance ensures that these latter give no contribution to the energy difference~\cite{carusotto}, whereas in the disordered case they have shown to develop a finite contribution, as emerged from numerical simulations. In Fig. \ref{fig_superfluid} the superfluid fraction is reported as a function of the interaction energy for three different fixed values of disorder. The computation of the twisted problem being quite demanding in terms of computational resources, thus limiting the number of disorder realizations, results in the large error bars in the plots. The effect of the finite size simulation is that the computed superfluid fraction is expected to be slightly larger than the real one and only a careful size-scaling analysis gives reliable quantitative information about the thermodynamic limit. To improve the averaging procedure we compute the energy of a system with twisted boundary conditions splitting it into $N$ bins as
\begin{equation}\label{E_theta}
E_\Theta=\sum_i^N \rho_{Si} (\theta_i-\theta_{i-1})^2,
\end{equation}
where $\rho_{Si}$ is the superfluid fraction of the $i^{th}$ bin and $\theta_i,\theta_{i-1}$ are the phases at the boundary of the $i^{th}$ cell. With the constraint
\begin{equation}
\sum_i \theta_i = \Theta,
\end{equation}
it can be shown that minimizing the energy (\ref{E_theta}) corresponds to taking the harmonic average of the superfluid fractions, i.e
\begin{equation}
\rho_S=\left(\sum_i \frac{1}{\rho_{Si}}\right)^{-1}.
\end{equation}
For this reason each point shown in Fig. \ref{fig_superfluid} is computed as a harmonic mean of the superfluid fraction of each realization and the error bars are computed accordingly.  The shaded zone shows the phase boundary predicted by studying the long-range decay of the one body density matrix.

In Fig. \ref{fig_superfluid} it is evident that the boundary computed by means of the correlation length coincides with the prediction based on superfluidity. In fact, the zero-value for the superfluid fraction is consistent with the error bars of all the cases belonging to the insulator phase, whereas the superfluid points acquire a finite $\rho_S$. It is worth noticing that the average procedure is most demanding when close to the boundary and is reflected in larger error bars in the proximity of the phase transition.

\section{Phase Diagram}\label{sec_pd}
With the methods explained so far we can characterize the phase of the Bose gas. We are able to draw the mean field phase diagram of the 1D Bose gas at zero temperature as a function of disorder and interaction energies.
\begin{figure}
\includegraphics[width=0.5\textwidth]{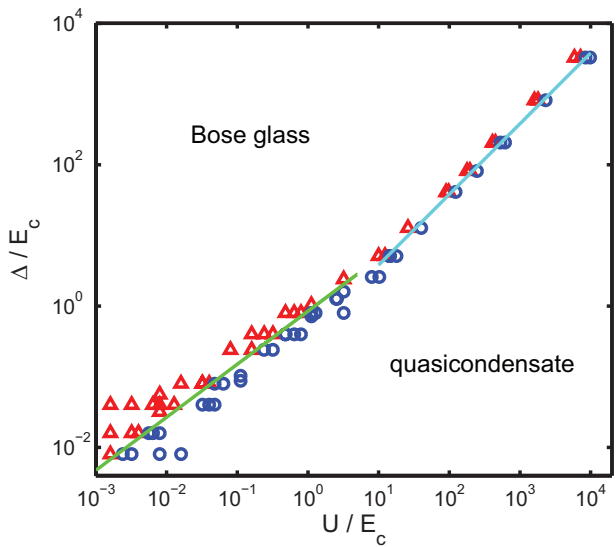}
\caption{(color online) Sketch of the phase diagram of the 1D Bose gas as a function of interaction and disorder. ($\triangle$): Bose glass; ($\bigcirc$): quasi-condensate.}
\label{fig_pd}
\end{figure}

We presented an earlier version of this phase diagram in Ref.~\cite{fontanesi}. In Fig. \ref{fig_pd} we show an extended phase diagram that includes the TF regime. It clearly shows two different trends depending on the ratio between the characteristic energies at the transition and $E_c$. These regimes can be identified by the ratio $\kappa=U/E_c$, in fact the limit $\kappa\ll 1$ represents the WN limit, where the healing length is much longer than the disorder correlation length. The opposite case $\kappa \gg 1$ marks the TF regime. The numerical results give two power-law dependencies of the boundary in these limiting cases $\Delta/E_c=C(U/E_c)^\gamma$, with $\gamma$ equal to $3/4$ and $1$ respectively. The lower part of the phase diagram represents the WN limit: in this regime a single energy scale characterizes the disordered potential~\cite{halperin} (see Appendix)
\begin{equation}
E_0=\Delta\left(\frac{\Delta}{E_c}\right)^{1/3}.
\end{equation}
Thus, assuming that at the transition the interaction energy is proportional to $E_0$, we directly obtain
\begin{equation}
\frac{U}{E_c}=C_1\left(\frac{\Delta}{E_c}\right)^{4/3},
\end{equation}
that correctly reproduces the power-law found numerically. In the opposite regime, i.e. the TF regime $\mu\gg E_c$, the scale of the potential is much larger than the typical length involved in the modulation of the ground state wavefunction ($\xi\ll\eta$). In this limit $E_c$ is no longer relevant and the thermodynamic phase is only determined by a one-to-one competition between disorder and interaction energies.
The computed phase diagram is valid in the weakly interacting limit and becomes exact in the limit $\rho\to \infty$ at constant $g\rho$. The phase diagram shows an infinite slope in the origin, in fact the power-law is smaller than $1$, and this is in agreement with previous theoretical calculations~\cite{falco}. This condition implies that in an experiment where the correlation length $\eta$ is reduced at constant interaction and disorder amplitudes, one would always end up in the superfluid phase. The obtained phase diagram is strikingly similar to the diagram obtained by investigating the change in the density profile of the $1D$ Bose gas~\cite{lugan_private}. A precise link between the phase transition and the fragmentation of the ground state wavefunction will be the object of a future study.

For the proportionality constant in the WN limit, we numerically find $C_1\sim1.1$. This is in good agreement with the prediction~\cite{altsh} that the proportionality between $E_0$ and $U$ should be approximately $1$, ($E_0/U\simeq 1$).

As stated above the mean-field description does not hold in the strongly interacting regime, that leads finally to an interaction dominated Bose glass phase. Consequently, the reentrant Bose glass phase obtained in the discrete model for strong interaction~\cite{giamarchi} cannot be described within this mean-field model. Moreover, the model cannot be applied in the disorder-dominated case, where the coherence extends only within a few maxima of the density and the system is in the so-called Lifshitz glass phase~\cite{lugan_phase}.

\section{Conclusions}\label{section_conclusions}

We have studied the phase diagram of a 1D Bose-gas at zero temperature in presence of correlated disorder. We analyzed the changes in the Bogoliubov excitations that entail the phase transition: we have found that the DOS diverges in the Bose-Glass phase while it approaches a constant value in the quasicondensed case. Moreover the localization of the excitations always shows an $E^{-\alpha}$ divergence and $\alpha=1$ marks the phase transition.
We have established the phase diagram by inspecting the long range decay of the one-body density matrix. This analysis led to the identification of two regimes in which the boundary follows a power-law relation between disorder and interaction: a WN zone, where a $3/4$ power-law relation holds, and a TF regime, where the relation becomes linear. This phase diagram has been confirmed by inspecting the superfluid fraction of the system.

This theoretical analysis could be very useful for future investigations aimed at the determination of the superfluid to Bose glass phase transition in 1D weakly interacting alcali gases.

\begin{acknowledgments}
We are grateful to T.~Giamarchi, S.~Giorgini, P.~Lugan and L.~Sanchez-Palencia for enlightening discussions. This work was supported by the Swiss National Science Foundation through project No. 200021-117919.
\end{acknowledgments}

\appendix{}
\section{White noise limit}

The WN limit argument applies to arbitrary dimensionality $D$ to predict the behaviour of the phase boundary at the origin. The correlation of the potential is given by~\cite{savona} $\langle V(\mathbf{r})V(\mathbf{r'})\rangle=\Delta^2 f_{\mathbf{r}-\mathbf{r'}}$, where for a gaussian correlation we have $f_{\mathbf{r}-\mathbf{r'}}=e^{-{(\mathbf{r}-\mathbf{r'})^2}/{2\eta^2}}$. In the WN case the potential is equivalent to a delta correlated one,
\begin{equation}\label{delta_WN}
\langle W(\mathbf{r})W(\mathbf{r'})\rangle=w\delta(\mathbf{r}-\mathbf{r'}),
\end{equation}
so that, if we assume $\hbar^2/2M=1$, $w$ has the dimensionality $[E^{2-D/2}]$ and in the WN limit every quantity having dimension of an energy must be proportional to $E_0=w^{2/(4-D)}$. On the other hand, if we take the limit for $\eta\to 0$ of the Gauss-correlated potential we get $f_{\mathbf{r}-\mathbf{r'}}\propto \eta^{D}\delta(\mathbf{r}-\mathbf{r'})$. Comparison with Eq. (\ref{delta_WN}) implies $w\sim \Delta^2 \eta^{D}$. In particular the critical interaction energy should be proportional to $E_0$ in the WN limit. Using $\eta^D\propto E_c^{-D/2}$, we conclude that
\begin{equation}
\frac{\Delta}{E_c}\propto\left(\frac{U}{E_c}\right)^{1-\frac{D}{4}}.
\end{equation}
The prediction for the $2D$ and $3D$ cases are respectively $\kappa=1/2,1/4$. Thus the slope in the origin remains infinite and the difference with respect to the linear relation in the TF limit becomes more pronounced in higher dimensions. These results are the same that have been found by Falco and coworkers~\cite{falco}.

\end{document}